\documentclass[aps,prb,twocolumn,showpacs,amsmath,amssymb,superscriptaddress, citeautoscript]{revtex4-2}
\setcitestyle{super}
\usepackage[cmtip,all]{xy}
\usepackage{graphicx}
\usepackage{hyperref}
\usepackage{xcolor}
\usepackage{color}
\usepackage[all,cmtip]{xy}
\usepackage{multirow}
\usepackage{natbib}
\usepackage{amsmath,amsthm} 
\usepackage{amsfonts}

\makeatother

\begin{document}

\title{Eightfold Degenerate Dirac Nodal Line in Collinear Antiferromagnet Mn$_{5}$Si$_{3}$}

\author{Victor Mendoza-Estrada}
\email{evictor@uninorte.edu.co}
\affiliation{Departamento de F\'{i}sica y Geociencias, Universidad del Norte, Km. 5 V\'{i}a Antigua Puerto Colombia, Barranquilla 081007, Colombia}
\affiliation{Facultad de Ciencias, Educación, Artes y Humanidades, Institución Universitaria de Barranquilla, Cra 45 No. 48-31, Barranquilla, Colombia}
\affiliation{Área de Ciencias Naturales, Colegio San José, Via 5 CUA-175, Puerto Colombia, Colombia}

\author{Rafael Gonz\'{a}lez-Hern\'{a}ndez}
\email{rhernandezj@uninorte.edu.co}
\affiliation{Departamento de F\'{i}sica y Geociencias, Universidad del Norte, Km. 5 V\'{i}a Antigua Puerto Colombia, Barranquilla 081007, Colombia}

\author{Bernardo Uribe}
\email{bjongbloed@uninorte.edu.co}
\affiliation{Departamento de Matem\'{a}ticas y Estad\'{i}stica, Universidad del Norte, Km. 5 V\'{i}a Antigua Puerto Colombia, Barranquilla 081007, Colombia}

\author{Libor ~\v{S}mejkal}
\email{lsmejkal@pks.mpg.de}
\affiliation{Max-Planck-Institut für Physik Complexer Systeme, Nöthnitzer St. 38, 01187, Dresden, Germany}
\affiliation{Max-Planck-Institut für Chemische Physik Fester Stoffe, Nöthnitzer St. 40, 01187, Dresden, Germany}
\affiliation{Institute of Physics, Academy of Science of the Czech Republic, Cukrovarnick\'a 10, 162 00 Praha 6, Czech Republic}

\date{\today}

\begin{abstract}
	We study the electronic, magnetic, and spin transport properties of the orthorhombic Mn$_{5}$Si$_{3}$ compound in the $AF2$ phase using symmetry analysis and $ab$-$initio$ calculations. 
	Our ground state energy calculations align with experimental observations, demonstrating that the collinear antiferromagnetic (AFM) order, with N\'{e}el vector in the [010] direction, is the most stable magnetic configuration both with and without spin-orbit coupling (SOC) in a bulk lattice geometry.  
	We identified an unconventional eight-fold degenerate Dirac nodal line (DNL) close to the Fermi level, characterized by negligible SOC. This DNL is robustly protected by a unique combination of  a pure-spin symmetry and a lattice symmetry together with magnetic space group symmetries.
	Upon introducing SOC, this degeneracy is reduced to two four-fold DNLs, being protected by the combination of time-reversal, partial translation and nonsymmorphic symmetries within the magnetic space group. 
	We predict also a large intrinsic spin Hall conductivity (SHC)  which correlates with the presence of SOC-induced splitting of these eight-fold degenerate DNLs near the Fermi level. 
	These intriguing characteristics position collinear antiferromagnet Mn$_{5}$Si$_{3}$ as a compelling candidate for spintronic applications, particularly in the generation and detection of spin currents, while remaining compatible with modern silicon technology.
\end{abstract}

\maketitle	

\section{Introduction}

The generation, manipulation, and detection of spin currents are key aspects of the field of spintronics  \cite{Sinova2015}. Therefore, since its discovery twenty years ago, the Spin Hall effect (SHE) has become a prominent focus as it generates spin currents from charge currents \cite{Kato2004,Wunderlich2005,Day2005}.
The SHE generally arises from relativistic spin-orbit coupling (SOC), which links the spin and orbital degrees of freedom. There are different mechanisms that contribute to the SHE due to the difference in types of spin-orbit coupling: two extrinsic mechanisms, 'Skew scattering' and 'Side jump,' are due to sample impurities, and an intrinsic mechanism that depends solely on the band structure and is proportional to the integration over the Fermi sea of the Berry curvature of each occupied band \cite{Sinova2015}. In materials with bands that have strong SOC, the intrinsic mechanism typically dominates the SHE \cite{Guo2008,Tanaka2008}.

Recent research by Derunova $et$ $al$. \cite{Derunova2019} has shown that in materials such as W$_{3}$Ta, Ta$_{3}$Sb, and Cr$_{3}$Ir, several symmetry operations combined with heavy atoms results in multiple Dirac crossings in the electronic band structure. 
In the absence of symmetry protection, SOC opens gaps at these crossings, generating a large intrinsic SHE. 
Similar effects were reported by Guo $et$ $al$. \cite{Guo2008}, who attributed the large intrinsic SHE in Pt to the SOC band splittings derived from doubly degenerate $d$ orbitals near high-symmetry points, such as L and X, close to the Fermi level.

The SHE has also been observed in antiferromagnets (AFs) \cite{Zhang2014,Zhang2016,Mendes2014}, but studies on the SHE in AF materials remain relatively limited \cite{Freimuth2010}.
In non-collinear AFMs, the SHE can arise even without SOC because the non-collinear magnetic order breaks spin-rotation symmetry, coupling spin to the lattice in a manner analogous to SOC \cite{Zhang2018,Gonzalez2021}. 
However, in collinear AFM systems, spin conservation prevents the intrinsic SHE in the absence of SOC, meaning that SOC is a necessary ingredient for the SHE to occur in these systems.

A breakthrough in the electrical control of antiferromagnetic moments has opened new opportunities for AFM materials in spintronic devices, much like ferromagnets \cite{Wadley2016,Schuler2016}. 
This has led to significant interest in the potential of AFM materials as active components in future spintronic technologies \cite{Jungwirth2016}. 
Among the various materials investigated for spintronic applications, intermetallic compounds containing manganese have attracted considerable attention due to their strong AFM interactions, abundance, and cost-effectiveness. 
Moreover, their compatibility with silicon—the foundational material of semiconductor technology—significantly enhances their appeal.
\cite{Chen2014,Nayak2016,Nakatsuji2015}
Consequently, the antiferromagnet Mn$_{5}$Si$_{3}$ has emerged as a particularly promising material within the scientific community, fostering extensive research over the past five decades \cite{Chen2014,Aronsson1960,Radovskii1965,Lander1967,Lander1967-2,Gel'd1975,Povzner1978,Menshikov1990,Vinokurova1990,Brown1992,A1-Kanani1995,Brown1995,Vinokurova1995,Irizawa2002,Ramos2002,Leciejewicz2008,Martins2009,Gottschilch2012,Surgers2014,Das2016,Surgers2016,Surgers2017,Bie2020}.

While the thin films of Mn$_5$Si$_3$ were recently discussed as candidates for altermangetism \cite{Reichlova2024,Reichlova2024arxiv,altermagnetism2}, the bulk compound Mn$_{5}$Si$_{3}$ has three magnetic phases: Antiferromagnetic 1 (AF1) with non-collinear magnetic moments order, Antiferromagnetic 2 (AF2) with collinear order, and Paramagnetic (PM) phase. 
The PM phase has a hexagonal structure at room temperature with space group $P6_{3}/mcm$ and lattice parameters $a$ = 6.910 \AA, $c$ = 4.814 \AA \cite{Aronsson1960}. 
The non-collinear AF1 magnetic structure is stable below 66 K and has monoclinic symmetry. In the AF2 phase, the dimensions of the orthorhombic cell at 70 K are $a$ = 6.89856 \AA, $b$ = 11.89120 \AA, and $c$ = 4.79330 \AA. 
The orthorhombic cell is face-centered with space group $Ccmm$, containing four molecules per unit cell (Z = 4) \cite{Brown1995}.

Previously, density functional theory (DFT) was used to investigate the density of states \cite{Gottschilch2012} and exchange constants \cite{Biniskos2018,Biniskos2022,Biniskos2023} of Mn$_{5}$Si$_{3}$  in different phases.  
Our study presents first-principles ($ab$-$initio$) calculations to elucidate the electronic and magnetic behavior of this compound in the collinear antiferromagnetic phase ($AF2$). 
The presence of Dirac nodal planes and eight-fold degenerate DNL is predicted in the absence of SOC.  
The incorporation of the pure-spin symmetry $ [\widetilde{C_{2y}}||\mathrm{E}] $ and the lattice symmetry $[\mathrm{E}||t_{0,0,\frac{1}{2}}M_z]$ into the magnetic space group symmetries that fixes the path U-R enforces the eight-fold degeneration.
Upon including SOC, the Dirac nodal planes and eight-fold degenerate DNL split, except along the Z-U path, where band degeneracy is protected by nonsymmorphic symmetries. 
The splitting of the eight-fold degenerate DNL near the Fermi level generates large values in the spin Berry curvature and, consequently, a large intrinsic spin Hall effect (SHE) in this collinear AFM.

\section{Electronic and magnetic properties}

\begin{figure*}[!htb]
	\center
	\includegraphics[scale=0.59]{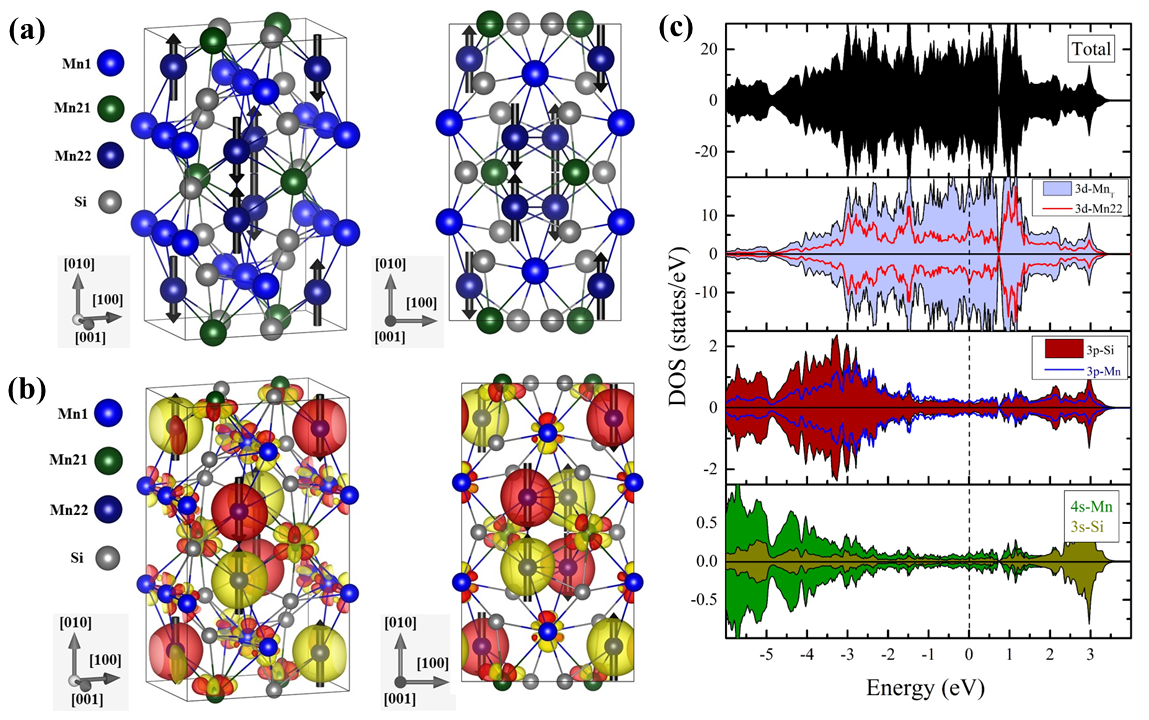}
	\caption{(a) Magnetic structure of the AF2 phase of Mn$_{5}$Si$_{3}$. The brown, purple, and red circles represent the inequivalent positions of Mn1, Mn21, and Mn22 atoms, respectively. The arrows indicate the direction of the magnetic moments of the Mn22 atoms. (b) Spin charge density ($\rho_{\uparrow} - \rho_{\downarrow}$) projected along the [010] direction of Mn$_{5}$Si$_{3}$. (c) Density of states (DOS) of Mn$_{5}$Si$_{3}$ in the AF2 phase without SOC.}
	\label{fig1}
\end{figure*}

Figure \ref{fig1}(a) shows the magnetic structure of the AF2 phase (orthorhombic unit cell) of the Mn$_{5}$Si$_{3}$ compound proposed by Brown $et$ $al$. \cite{Brown1992}, where the magnetic moments (represented by black arrows) are exclusively located at the Mn22 atom sites. To study the magnetic order of this phase, total energy calculations were initially performed with and without spin-orbit coupling (SOC) for the non-magnetic (NM), ferromagnetic (FM), and collinear antiferromagnetic (AF2) states of the orthorhombic Mn$_{5}$Si$_{3}$ compound. The spin directions were defined with respect to the coordinate axes ($x,y,z$) (corresponding to the crystallographic directions [100], [010], and [001], respectively). The energy differences $\Delta E$=$E_{FM}$-$E_{AFM}$ indicate a high stability of the AF2 state compared to the FM state along the [010] direction, approximately 79.3 and 229.8 $m$eV per formula unit of Mn$_{5}$Si$_{3}$ without and with SOC, respectively. 

In the absence of SOC, the energy of the antiferromagnetic AF2 state is degenerate with respect to spin orientation. However, the introduction of SOC introduces a magnetic anisotropy that differentiates between the minimum energy orientation (easy axis) and the maximum energy orientation (hard axis). As shown in Table \ref{table1}, SOC induces a magnetic anisotropy with the easy axis oriented along the [010] direction, thereby establishing it as the preferred magnetic orientation in the AF2 state of Mn$_{5}$Si$_{3}$. This behavior is in good agreement with theoretical results \cite{Dos-Santos2021} and experimental findings \cite{Brown1992}, where it has been demonstrated that between temperatures of 60 K and 100 K, the Mn$_{5}$Si$_{3}$ compound exhibits antiferromagnetic behavior with Mn22 magnetic moments aligned parallel and antiparallel to the crystallographic $b$-axis in a collinear arrangement.

\begin{table}[h]
	\begin{center}
		\caption{Total Energy ($E_{AF2}$) in eV, Magnetic Anisotropy Energy (MAE) in meV, Average Magnetic Moment ($\overline{M}$) in $\mu_B$ per Mn atom, and Total Magnetic Moment ($M_{T}$) in $\mu_B$ of Mn$_{5}$Si$_{3}$ in the AF2 phase for N\'{e}el vector orientations.}
		\label{table1}
		\vspace{0.1cm}
		\scalebox{0.9}{
			\begin{tabular}{lccc}
				\hline \hline
				\textbf{Property} & \textbf{[100]} & \textbf{[010]} & \textbf{[001]} \\
				\hline
				{Tot. Energy} & -255.678 & -255.747 & -255.680 \\
				{MAE (meV)} & 17.40 & 0.00 & 16.80 \\
				{$\overline{M}$(Calc)}& 2.63, 0.01, 0.00 & 0.02, 2.64, 0.00 & 0.00, 0.00, 2.63 \\
				{$\overline{M}$(Theo)} & --- & 0.00, 2.54, 0.00 & --- \\
				{$\overline{M}$ (Exp)} & --- & 0.26, 1.87, 0.00 & --- \\
				{Tot. $M_{T}$} & 0.29 & 0.00 & 0.02 \\
				\hline \hline
		\end{tabular}}
	\end{center}
	\vspace{0.05cm}
\end{table}

Table \ref{table1} presents the total magnetic moment ($M_{T}$) and the average local magnetic moment ($\overline{M}$) of Mn${22}$ atoms in Mn$_{5}$Si$_{3}$ for the [100], [010], and [001] directions. The table shows non-zero $M{T}$ values for the [100] and [001] directions, indicating incomplete cancellation between antiferromagnetic and ferromagnetic regions. The average local magnetic moments of Mn22 atoms are similar across all directions, reflecting minimal dependence on the magnetic spin configuration.

In Figure \ref{fig1}(b), the spin charge density of the Mn$_{5}$Si$_{3}$ compound is shown, with calculations performed along the [010] direction, which is the preferential easy axis. It is evident that the Mn22 atoms present the largest contribution to the local magnetic moments, where the centers of the atoms lie in highly symmetric regions of positive spin charge density $\rho_{\uparrow}$ (yellow) and negative spin charge density $\rho_{\downarrow}$ (red), forming zones of completely compensated spin charge density. On the other hand, the Mn1 and Mn21 atoms exhibit small contributions of $\rho_{\uparrow}$ and $\rho_{\downarrow}$ (yellow-red zone) associated with their atomic orbitals, which cancel each other out and could explain the absence of magnetic moments at these sites. These contributions of $\rho_{\uparrow}$ and $\rho_{\downarrow}$ could be attributed to the presence of the local field produced by the antiparallel magnetic moments of neighboring Mn22 atoms.

The calculated average magnetic moment of the Mn22 atoms is around 2.636 $\mu_{B}$ with a small inclination of $\sim$ 0.4$^\circ$ with respect to the crystallographic direction [010] (see Table \ref{table1}). This result is higher than the experimentally reported values of magnetic moments at 80 K and 70 K with inclinations of 3$^\circ$ and 8$^\circ$, respectively, which are 1.746 $\mu_{B}$ and 1.890 $\mu_{B}$ \cite{Gottschilch2012}. However, it is in good agreement with the theoretical results reported in the literature (2.400 $\mu_{B}$ and 2.544 $\mu_{B}$) \cite{Dos-Santos2021,Bradlyn2017,Vergniory2019,Vergniory2022,Topological,Bilbao}. The small differences in the calculated magnetic moments could depend on the functional used in the DFT calculations. Marcel Bornemann $et al$. \cite{Bornemann2019} showed that the magnetic moment of Mn in the B20-MnGe compound differs when using the LDA and PBEsol functionals. 

In Figure \ref{fig1}(c), the total density of states (TDOS) and partial density of states (PDOS) of 3d-Mn1, 3d-Mn21, 3d-Mn22, 3p-Si, 3p-Mn, 3s-Si, and 3s-Mn orbitals of the most stable AF2 phase along the [010] direction without SOC are shown. In this figure, it can be observed that the Mn$_{5}$Si$_{3}$ compound exhibits a metallic behavior, where energy bands around the Fermi level are mainly formed by the orbitals of the 3d-Mn states, while the Si states do not significantly contribute to the density of states at and near the Fermi level. The TDOS shows a symmetric distribution between the majority and minority spin regions, indicating the antiferromagnetic behavior of the system (see Table \ref{table1}).
The features present in the PDOS suggest a strong hybridization between the 3$d$-Mn and 3$p$-Si orbitals in the middle region of the valence band, similar results were reported for Mn$_{5}$Si$_{3}$ in the paramagnetic phase \cite{Vinokurova1990}. Since the distances between the Mn atoms vary, and the distance between pairs of Mn and Si atoms in the crystal structure changes, the hybridization between different types of Mn atoms for certain pairs (Mn-Si) is modified. This is because the bands of the Mn atoms are highly sensitive to their local environment, which could be the reason for the changes found in the interactions presented in the DOS. On the other hand, due to the short distances between the Mn1, Mn21, and Mn22 atoms, the interactions between the 3$d$ orbitals of these atoms near the Fermi level are very strong. Therefore, the mechanism of direct $d$-$d$ exchange between these atoms could be associated with the antiferromagnetism present in Mn22.

\begin{figure*}[!htb]
	\center
	\includegraphics[scale=0.62]{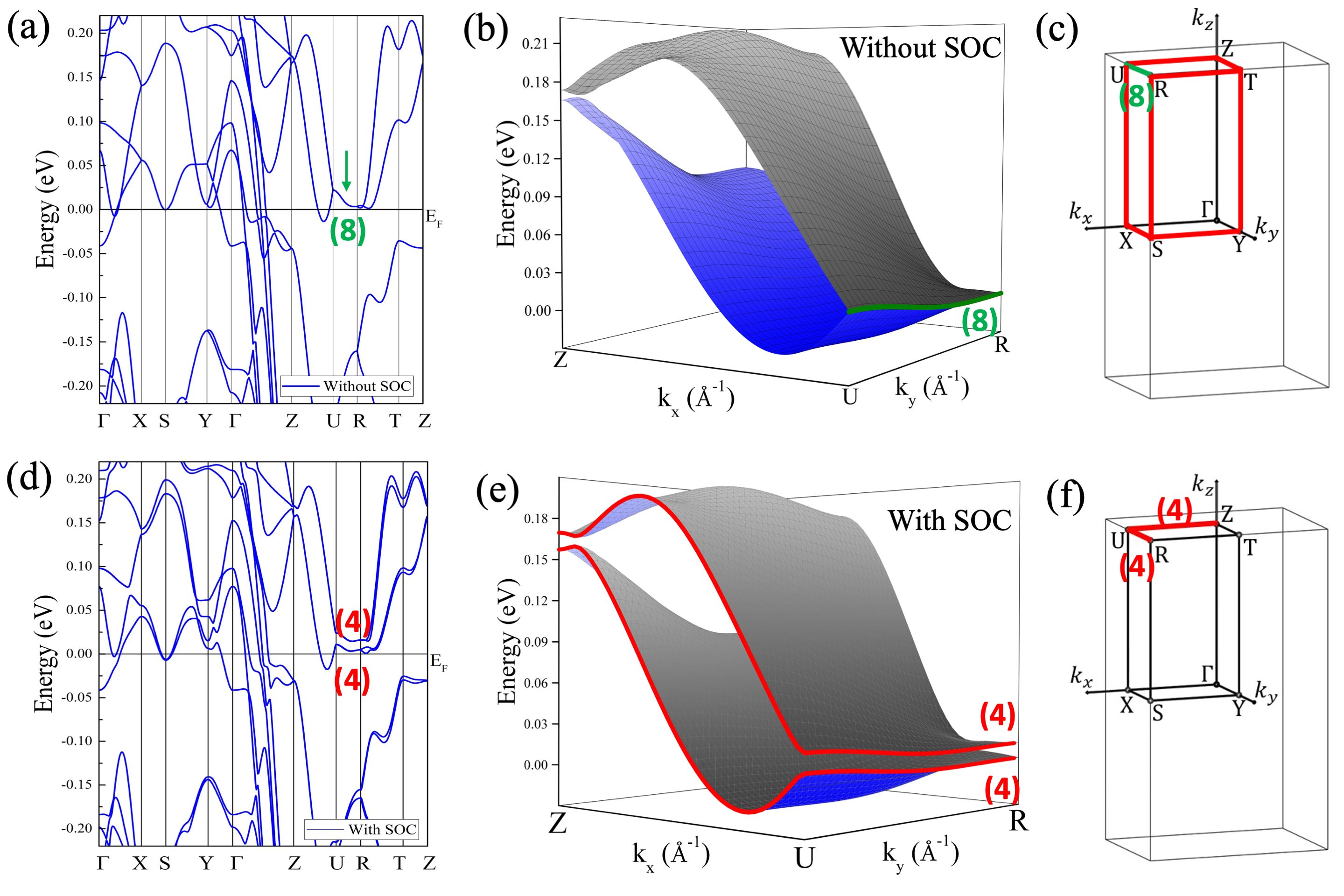}
	\caption{(a) Band structure of Mn$_{5}$Si$_{3}$ in the AFM2 phase without SOC. (b) 3D view of the eight-fold degenerate DNL along the U-R path (green line) in the $k_{z} = \pi$ plane without SOC. (c) First Brillouin zone (BZ) with the representation of the DNL (red lines) and the eight-fold degenerate band (green line) without SOC. (d) Band structure of Mn$_{5}$Si$_{3}$ in the AFM2 phase with SOC. (e) 3D view of the DNLs (red lines) along the Z-U and U-R path in the $k_{z} = \pi$ plane with SOC. (f) First Brillouin zone with the representation of the DNLs (red line) along the Z-U and U-R path with SOC. Band degeneracy is indicated in parentheses.}
	\label{fig2}
\end{figure*}

\section{Band structure and Symmetry analysis}

In the absence of SOC (Figure \ref{fig2}(a)), the band structure reveals that the spin-up and spin-down bands exhibit antiferromagnetic Kramers degeneracy \cite{PhysRevLett.118.106402,altermagnetism2}. 
This degeneracy is fundamentally protected by specific symmetry operations, which play a critical role in preserving the characteristics of the band structure \cite{Topp2016,Takahashi2017,Fang2015,Fu2018}.
In this regime, the spatial and spin degrees of freedom are partially unlocked, suggesting the existence of an pure spin symmetry, denoted as R$_{S}$, that inverts the spin direction while preserving the invariance of the wave vector \cite{altermagnetism1,McClarty2022,McClarty2024,McClarty2024a}. 
The symmetry operations that leave the magnetic crystal invariant in this framework are referred to as spin space operations, which are part of the spin space groups (SSG) \cite{altermagnetism1,SSG-classification,SSG-1977,SSG-2022}.
Recently, the study of SSG has gained considerable attention, particularly with the emergence of the novel altermagnetic phase \cite{altermagnetism1,altermagnetism2}. We include in the Appendix
a summary of the properties of the SSG and their relation with the magnetic space groups (MSG); we also choose a preferred basis in which both types of symmetries are parameterized.

When the SOC is included (Figure \ref{fig2}(d)), we identified several symmetry operations that reproduce the crystal structure of the collinear AFM phase of Mn$_{5}$Si$_{3}$. 
In this context, these symmetries are associated with the well-established magnetic space groups (MSG) \cite{MSG-book}.
The unitary MSG symmetries relevant to AF2 phase of Mn$_{5}$Si$_{3}$ are as follows:


\begin{align} 
	& \mathrm{E}, \widetilde{P},    t_{0,0,\frac{1}{2}} \widetilde{C_{2x}},  t_{0,0,\frac{1}{2}} \widetilde{M_x},\\
	 & t_{\frac{1}{2},\frac{1}{2},0}  \widetilde{C_{2y}},t_{\frac{1}{2},\frac{1}{2},0}  \widetilde{M_y},  t_{\frac{1}{2},\frac{1}{2},\frac{1}{2}} \widetilde{C_{2z}}, t_{\frac{1}{2},\frac{1}{2},\frac{1}{2}} \widetilde{M_z},
\end{align}
plus the antiunitary ones
\begin{align}
	&t_{\frac{1}{2},\frac{1}{2},0} \mathbb{T}   , t_{\frac{1}{2},\frac{1}{2},0} \mathbb{T}  \widetilde{P},  t_{\frac{1}{2},\frac{1}{2},\frac{1}{2}} \mathbb{T}  \widetilde{C_{2x}},    t_{\frac{1}{2},\frac{1}{2},\frac{1}{2}} \mathbb{T} \widetilde{M_x},\\
	& \mathbb{T}  \widetilde{C_{2y}}, \mathbb{T}\widetilde{M_y},  t_{0,0,\frac{1}{2}} \mathbb{T} \widetilde{C_{2z}}, t_{0,0,\frac{1}{2}} \mathbb{T} \widetilde{M_z}.
\end{align}

Here $\mathbb{T} $ denotes the time reversal operator with  $\mathbb{T}^2=-1 $, $t$ stands for  translation, $P$ denotes the inversion operator, $C_{2z}$ the 2-fold rotation around the $z$-axis, $M_z$ the mirror reflection on the $z$-plane, and the tilde over the
geometrical symmetries means that they are understood as operators; for instance, the rotation satifies $({C_{2z}})^2=1$ while the operator  satisfies $(\widetilde{C_{2z}})^2=-1$ due to SOC.


When the N\'{e}el vector is parallel to the $y$-axis the SSG symmetries includes
\begin{align}
	[\mathbb{T} \widetilde{C_{2x}}||\mathrm{E}], [\mathbb{T} \widetilde{C_{2z}}||\mathrm{E}], [ \widetilde{C_{2y}}||\mathrm{E}] 
\end{align}
and their composition with the MSG symmetries. 
In the SSG notation, the left hand side of the symmetry operations corresponds only to spin  symmetries while the right hand side corresponds to lattice symmetries (see the Appendix for more details).

\subsubsection{Magnetic Space Group: SOC included}

\paragraph{U-Z path.}
To investigate the band structure across different regions of the Brillouin zone (BZ), we start on the line defined by ($k_y$ = 0) and ($k_z$ =$ \pi$), corresponding to the U-Z path in the BZ (Figure \ref{fig2}(f)). 
Along this symmetry line, the eigenstates of the Hamiltonian display a four-fold degeneracy.

To analyze this degeneracy, we consider the following MSG operators:

\begin{align}
	\widehat{M_z}: &= t_{\frac{1}{2},\frac{1}{2},\frac{1}{2}}  \widetilde{M_z}\\
	\mathbb{T}\widehat{P}&:= t_{\frac{1}{2},\frac{1}{2},0} \mathbb{T} \widetilde{P} \\
	\mathbb{T} \widehat{M_x}&:=t_{\frac{1}{2},\frac{1}{2},\frac{1}{2}} \mathbb{T} \widetilde{M_x}
\end{align}
satisfying 
\begin{align}
	\widehat{M_z}^2 &= -e^{-ik_x-ik_y}|_{k_y=0}= -e^{-ik_x}\\
	(\mathbb{T}\widehat{P})^2 &= -1 \\
	(\mathbb{T}\widehat{M_x})^2 &= -e^{-ik_y-ik_z}|_{k_y=0, k_z=\pi}= 1
\end{align}
\begin{align}
	t_{1,1,1}(\mathbb{T} \widehat{P}) \widehat{M_z}&=\widehat{M_z} (\mathbb{T} \widehat{P})\\
	e^{-ik_x-ik_y-ik_z}|_{k_y=0, k_z=\pi} (\mathbb{T} \widehat{P}) \widehat{M_z}&=\widehat{M_z} (\mathbb{T} \widehat{P}) \\
	-e^{-ik_x} (\mathbb{T} \widehat{P}) \widehat{M_z}&=\widehat{M_z} (\mathbb{T} \widehat{P}) 
\end{align}
\begin{align}
	t_{1,0,0}(\mathbb{T} \widehat{M_x}) \widehat{M_z}&=t_{0,0,1}\widehat{M_z} (\mathbb{T} \widehat{M_x})\\
	-e^{-ik_x}(\mathbb{T} \widehat{M_x}) \widehat{M_z}&=e^{-ik_z}|_{ k_z=\pi} \widehat{M_z} (\mathbb{T} \widehat{M_x}) \\
	e^{-ik_x} (\mathbb{T} \widehat{M_x}) \widehat{M_z}&=\widehat{M_z} (\mathbb{T} \widehat{M_x}).
\end{align}

Now, take an eigenstate of the Hamiltonian which is also an eigenstate of $\widehat{M_z}$:
\begin{align}
	\widehat{M_z} \psi = i e^{-i\frac{k_x}{2} }\psi.
\end{align}


Since $\mathbb{T}\widehat{P}$ is antiunitary, it squares to $-1$, and leaves the momentum coordinates fixed, we have the equalities:
\begin{align}
	\langle \mathbb{T}\widehat{P} \psi| \psi \rangle = \overline{ \langle {(\mathbb{T}\widehat{P})}^2 \psi| \mathbb{T}\widehat{P}\psi \rangle }
	= \overline{ \langle- \psi| \mathbb{T}\widehat{P}\psi \rangle } = - \langle \mathbb{T}\widehat{P}\psi| \psi \rangle,
\end{align}
where the first equality follows from the antiunitary structure of the operator $\mathbb{T}\widehat{P}$. Recall that
any antiunitary operator can be written as the composition $U \mathbb{K}$ where $U$ is a unitary operator and  
$\mathbb{K}$ is complex conjugation. Hence $\langle U \mathbb{K} \psi |  U \mathbb{K} \phi  \rangle = 
\langle \mathbb{K} \psi |  \mathbb{K} \phi  \rangle = \overline{\langle \psi |  \phi  \rangle }$.

Therefore we have that $\langle \mathbb{T}\widehat{P} \psi| \psi \rangle=0$ and we can conclude that
 $\mathbb{T}\widehat{P}\psi$ and $\psi$ are linearly independent states. 


By the commutativity relations between $\mathbb{T}\widehat{P}$ and $\widehat{M_z}$ we have:
\begin{align}
	\widehat{M_z} (\mathbb{T}\widehat{P}) \psi &= -e^{-ik_x} (\mathbb{T}\widehat{P}) \widehat{M_z} \psi \\
	&= -e^{-ik_x} (\mathbb{T}\widehat{P}) (ie^{-i\frac{k_x}{2}} \psi)\\
	&=-e^{-ik_x}(-ie^{i\frac{k_x}{2}} ) (\mathbb{T}\widehat{P})  \psi \\
	&=ie^{-i\frac{k_x}{2}}  (\mathbb{T}\widehat{P})  \psi,
\end{align}
therefore both $\mathbb{T}\widehat{P}\psi$ and $\psi$ have the same $\widehat{M_z}$- eigenvalues.

Construct the eigenstate $\mathbb{T}\widehat{M_x} \psi$ and note that:
\begin{align}
	\widehat{M_z}(\mathbb{T}\widehat{M_x}) \psi & = e^{-ik_x}(\mathbb{T}\widehat{M_x}) \widehat{M_z} \psi \\
	& = e^{-ik_x}(\mathbb{T}\widehat{M_x}) (ie^{-i\frac{k_x}{2}} \psi )\\
	& = -ie^{-i\frac{k_x}{2}} (\mathbb{T}\widehat{M_x}) \psi,
\end{align}
therefore $\mathbb{T}\widehat{M_x} \psi$ has different $\widehat{M_z}$-eigenvalue
as $\mathbb{T}\widehat{P}\psi$ and $\psi$, and hence it is linearly independent of the two.
Constructing $(\mathbb{T}\widehat{M_x} )\mathbb{T}\widehat{P}\psi$ we get another
eigenstate with same $\widehat{M_z}$-eigenvalue as $\mathbb{T}\widehat{M_x} \psi$,
but also linearly independent of other three vectors.

Therefore, if $\psi$ is an eigenstate of the Hamiltonian and of $\widehat{M_z}$, the set of
eigenstates
\begin{align}
	\{ \psi, \mathbb{T}\widehat{P}\psi, \mathbb{T}\widehat{M_x} \psi, (\mathbb{T}\widehat{M_x} )\mathbb{T}\widehat{P}\psi \}
\end{align} 
define a four-fold degenerate DNL set of eigenstates localized on the line $k_y=0,k_z=\pi$ (U-Z path) as illustrated in Figure \ref{fig2}(e).

\paragraph{U-R path.}

Let us show also that along the line $k_x=\pi$, $k_z= \pi$ (U-R path) the eigenstates of the Hamiltonian are four-fold degenerate (Figure \ref{fig2}(f)). 
This symmetry line is also described using the MSG framework.

Consider the symmetries leaving the U-R line fixed:
\begin{align} 
	\mathbb{T}\widehat{P}:= t_{\frac{1}{2},\frac{1}{2},0} \mathbb{T} \widetilde{P}\\
	\widehat{M_x}=t_{0,0,\frac{1}{2}} \widetilde{M_x} \\
	\widehat{C_{2y}}= t_{\frac{1}{2}\frac{1}{2},0} \widetilde{C_{2y}}.
\end{align}
These symmetries satisfy:
\begin{align}
	(\mathbb{T}\widehat{P})^2=-1, \ \  \widehat{M_x}^2& =1 \ \ \widehat{C_{2y}}^2=e^{-ik_y}\\
	\widehat{M_x}\widehat{C_{2y}}&=-\widehat{C_{2y}}\widehat{M_x} \\ 
	\widehat{M_x}(\mathbb{T}\widehat{P})&=(\mathbb{T}\widehat{P})\widehat{M_x} \\
	e^{-ik_y}(\mathbb{T}\widehat{P})\widehat{C_{2y}}&=\widehat{C_{2y}}(\mathbb{T}\widehat{P})
\end{align}
The smallest matrix representation of $\widehat{M_x}$ and $\widehat{C_{2y}}$ satisfying the identities is:
\begin{align}
	\widehat{M_x} =  \left(\begin{matrix} 1& 0\\ 0  & -1
	\end{matrix} \right) \ \
	\ \widehat{C_{2y}} = e^{i\frac{k_y}{2}}\left(\begin{matrix} 0& 1\\ 1  & 0
	\end{matrix} \right).
\end{align}
Since $\langle (\mathbb{T}\widehat{P}) \psi | \psi \rangle $ is always zero, we know that
$(\mathbb{T}\widehat{P}) \psi $ has same $\widehat{M_x}$-eigenvalue as $\psi$. Therefore, if $\psi$ is 
an eigenvector of $\widehat{M_x}$ the following four states have the same energy:
\begin{align}
	\{ \psi, \widehat{C_{2y}}\psi,  (\mathbb{T}\widehat{P})\psi,  (\mathbb{T}\widehat{P})\widehat{C_{2y}}\psi \}.
\end{align} 

These four-fold degenerate DNLs are in agreement with the $ab$-$initio$ calculations presented in Figure \ref{fig2}(e).

\subsubsection{Spin Space Group: SOC not included}

\paragraph{$k_z=\pi$ plane.}
Switching SOC off, let us show that on the planes $k_z=\pi$ the energy is also four-fold degenerate, indicating the presence of a Dirac nodal plane.

Now, consider the operators in the SSG notation as explained in the Appendix:
\begin{align} 
	\mathbb{T}\widehat{P}:= t_{\frac{1}{2},\frac{1}{2},0} \mathbb{T} \widetilde{P} =[\mathbb{T}|| t_{\frac{1}{2},\frac{1}{2},0} P]\\ 
	\widehat{M_x}=t_{0,0,\frac{1}{2}} \widetilde{M_x} =[\widetilde{C_{2x}}|| t_{0,0,\frac{1}{2}} M_x]
\end{align}
together with the pure-spin symmetry
\begin{align}
	\mathcal{F}_y := [\widetilde{C_{2y}}||\mathrm{E}].
\end{align}
We have the following identities:
\begin{align} 
	( \mathbb{T}\widehat{P})^2=-1, \ \
	\widehat{M_x}^2=1,\ \ {\mathcal{F}_y}^2=-1
\end{align}
and commutation relations:
\begin{align}
	( \mathbb{T}\widehat{P}) \mathcal{F}_y = \mathcal{F}_y ( \mathbb{T}\widehat{P})\\
	\widehat{M_x} \mathcal{F}_y =-  \mathcal{F}_y \widehat{M_x}\\
	( \mathbb{T}\widehat{P}) \widehat{M_x} = \widehat{M_x}( \mathbb{T}\widehat{P}).
\end{align}

The operator $\widehat{M_x}$ has real eigenvalues $\pm 1$, and we may take the eigenvector $\psi$ satisfying $\widehat{M_x} \psi = \psi$. The eigenvector $( \mathbb{T}\widehat{P}) \psi$ 
has same $\widehat{M_x}$ eigenvalue as $\psi$ and it is different from $\psi$ because $( \mathbb{T}\widehat{P}) $
squares to $-1$ and it is antiunitary. 
Take the eigenstates $\mathcal{F}_y\psi$ and $\mathcal{F}_y ( \mathbb{T}\widehat{P})  \psi$, and 
note that both have opposite sign $\widehat{M_x}$-eigenvalue of $\psi$.
Therefore, if $\psi$ is an eigenstate of the operator $\widehat{M_x}$, the set of eigenstates
\begin{align}
	\{ \psi, (\mathbb{T}\widehat{P})\psi, \mathcal{F}_y\psi, \mathcal{F}_y(\mathbb{T}\widehat{P})\psi \}
\end{align}
define a four-fold degenerate set of eigenstates localized on the $k_z=\pi$, which defines the Dirac nodal plane.
This result is consistent with the electronic band structure presented in Figure \ref{fig2}(a).

A matrix representation for these symmetries is the following:
\begin{align}
	\widehat{M_x} =  \left(\begin{matrix} 
		0&1&0&0\\
		1&0&0&0\\
		0&0&0&1\\
		0&0&1&0
	\end{matrix} \right)  \ \ 
	\mathcal{F}_y =  \left(\begin{matrix} 
		i&0&0&0\\
		0&-i&0&0\\
		0&0&-i&0\\
		0&0&0&i
	\end{matrix} \right)  \\
	\mathbb{T}\widehat{P} = \left(\begin{matrix} 
		0&0&1&0\\
		0&0&0&1\\
		-1&0&0&0\\
		0&-1&0&0
	\end{matrix} \right) \mathbb{K}.
\end{align}

\paragraph{U-R path.}

In the absence of SOC, we find that along the line $k_x=\pi = k_z$ (U-R path), the eigenstates exhibit an unprecedented eight-fold degeneracy.
This finding represents unique characteristic that has not been previously reported in collinear AFM systems.

Localizing further on the line $k_x=\pi = k_z$, (U-R path in the Figure \ref{fig2}(c)), one may add the operator 
\begin{align} 
	\widehat{C_{2y}}= t_{\frac{1}{2}\frac{1}{2},0} \widetilde{C_{2y}} = [ \widetilde{C_{2y}}||t_{\frac{1}{2}\frac{1}{2},0}C_{2y}].
\end{align}
For this operator we have the following identities:
\begin{align}
	\widehat{C_{2y}}^2&=e^{-ik_y},  \\  \widehat{C_{2y}} \mathcal{F}_y &= \mathcal{F}_y  \widehat{C_{2y}} \\
	\widehat{C_{2y}}\widehat{M_x} &=-\widehat{M_x}\widehat{C_{2y}} \\
	e^{-ik_y}( \mathbb{T}\widehat{P})\widehat{C_{2y}} &=\widehat{C_{2y}}( \mathbb{T}\widehat{P})
\end{align}
and we have that the following matrix representation of the operators match all the previous identities:
\begin{align}
	\widehat{M_x} =  \left(\begin{matrix} 
		0&1&0&0\\
		1&0&0&0\\
		0&0&0&1\\
		0&0&1&0
	\end{matrix} \right)  \ \  
	\widehat{C_{2y}} = e^{-i\frac{k_y}{2}}\left(\begin{matrix} 
		1&0&0&0\\
		0&-1&0&0\\
		0&0&1&0\\
		0&0&0&-1
	\end{matrix} \right)\\
	\mathcal{F}_y =  \left(\begin{matrix} 
		i&0&0&0\\
		0&-i&0&0\\
		0&0&-i&0\\
		0&0&0&i
	\end{matrix} \right)  \ \
	\mathbb{T}\widehat{P} = \left(\begin{matrix} 
		0&0&1&0\\
		0&0&0&1\\
		-1&0&0&0\\
		0&-1&0&0
	\end{matrix} \right) \mathbb{K}.
\end{align}


Now, an interesting feature appears  whenever we try to add the following SSG operator, which
is a composition of a pure-spin operator with a MSG operator:


\begin{align}
	L_z:=[\mathbb{T}\widetilde{C_{2z}}||\mathrm{E}] \ [\mathbb{T}\widetilde{C_{2z}}||t_{0,0,\frac{1}{2}}M_z] = [\mathrm{E}||t_{0,0,\frac{1}{2}}M_z].
\end{align}

For this operator we have the identities
\begin{align}
	{L_z}^2&=1,  \\  L_z\mathcal{F}_y &= \mathcal{F}_y  L_z \\
	\widehat{C_{2y}} L_z &=-L_z\widehat{C_{2y}} \\
	\widehat{M_x} L_z &=-L_z\widehat{M_x} \\
	( \mathbb{T}\widehat{P})L_z &=-L_z( \mathbb{T}\widehat{P})
\end{align}

The following matrix representation for $L_z$ is the only matrix representation
that satisfies the first four of the previous five identities (besides a sign change)
\begin{align}
	L_z = \left(\begin{matrix} 
		0&0&0&1\\
		0&0&-1&0\\
		0&-1&0&0\\
		1&0&0&0
	\end{matrix} \right).
\end{align}
This means that the previous matrices for the operators $\widehat{M_x}, \mathcal{F}_y, \widehat{C_{2y}},
L_z$ define an irreducible representation of the group they generate. Note that all the operations are unitary.

When one checks the equation $( \mathbb{T}\widehat{P})L_z =-L_z( \mathbb{T}\widehat{P})$ with the 
prescribed matrices, one notes that the equation is not satisfied.  The matrices for $L_z$ and 
$( \mathbb{T}\widehat{P})$ commute. This implies that on a 4-dimensional vector space
the whole set of operators cannot be represented.

Following Wigner's criterion for the classification of irreducible corepresentations we need to determine 
whether the representation of the conjugate operators 
\begin{align}
	S \mapsto ( \mathbb{T}\widehat{P}) S ( \mathbb{T}\widehat{P})^{-1}
\end{align}
define an isomorphic representation to the original one  \color{black} (see page 340 of Wigner's book \cite{Wigner}). \color{black}

If we denote by $\Delta(S)$ the matrix representation of the unitary operators, the conjugate one is:
\begin{align}
	\bar{\Delta}(S)= \Delta(( \mathbb{T}\widehat{P}) S ( \mathbb{T}\widehat{P})^{-1})^*.
\end{align}
Therefore we have that 
\begin{align}
	\bar{\Delta}(\widehat{M_x})&=\Delta(\widehat{M_x})\\
	\bar{\Delta}(e^{i\frac{k_y}{2}}\widehat{C_{2y}})&={\Delta}(e^{i\frac{k_y}{2}}\widehat{C_{2y}})\\
	\bar{\Delta}(\mathcal{F}_y)&=-\Delta(\mathcal{F}_y)\\
	\bar{\Delta}(L_z)&=-\Delta(L_z).
\end{align}

The two complex representations $\Delta$ and $\bar{\Delta}$ are isomorphic with isomorphism
\begin{align}
	\beta:=   \left(\begin{matrix} 
		0&0&1&0\\
		0&0&0&1\\
		1&0&0&0\\
		0&1&0&0
	\end{matrix} \right)
\end{align}
namely, $\beta^{-1} \Delta \beta= \bar{\Delta}$.

In Wigner's criterion  \color{black}  (see page 341 of Wigner's book \cite{Wigner}), \color{black} 
when the two representations are isomorphic there are two situations:
\begin{align}
	\beta \beta^*=\Delta(a_0^2)\\
	\beta \beta^*=-\Delta(a_0^2)
\end{align}
In our case, the antiunitary element $a_0$ that we took is $a_0=( \mathbb{T}\widehat{P})$, and therefore
$a_0^2=( \mathbb{T}\widehat{P})^2=-1$. Since the equation that is satisfied is the second one,
in our case $\beta \beta^*=1$ and $-\Delta(a_0^2)=1$, then the corepresentation is of quaternionic type
and  the irreducible corepresentation consists of $\Delta \oplus \bar{\Delta}$. 
According to Wigner the correpresentation can be defined by the matrices:
\begin{align}
	u \mapsto   \left(\begin{matrix} 
		\Delta(u)&0\\
		0&\Delta(u)
	\end{matrix} \right)
\end{align}
for unitary operators, and
\begin{align}
	a\mapsto   \left(\begin{matrix} 
		0&\Delta(aa_0^{-1})\beta\\
		-\Delta(aa_0^{-1})\beta&0
	\end{matrix} \right) \mathbb{K}
\end{align}
for antiuntitary ones $a$ \color{black} (see page 343 of Wigner's book \cite{Wigner}).  \color{black}
Recall that $\mathbb{K}$ denotes complex conjugation.

Therefore along the path $k_x=\pi=k_z$ (U-R path) the energy band degeneracy is of dimension $8$, which is corroborated by the $ab$-$initio$ calculations shown in Figure \ref{fig2}(b), where the DNL is indicated by the green line. This remarkable feature is due to the presence of the magnetic symmetries that
preserves the U-R path, together with the pure-spin symmetry $ [\widetilde{C_{2y}}||\mathrm{E}] $ and the lattice symmetry
$[\mathrm{E}||t_{0,0,\frac{1}{2}}M_z]$. The incorporation of these two last symmetries 
to the magnetic ones forces the irreducible corepresentation to be of dimension 8.

\begin{figure*}[!htb]
	\center
	\includegraphics[scale=0.6]{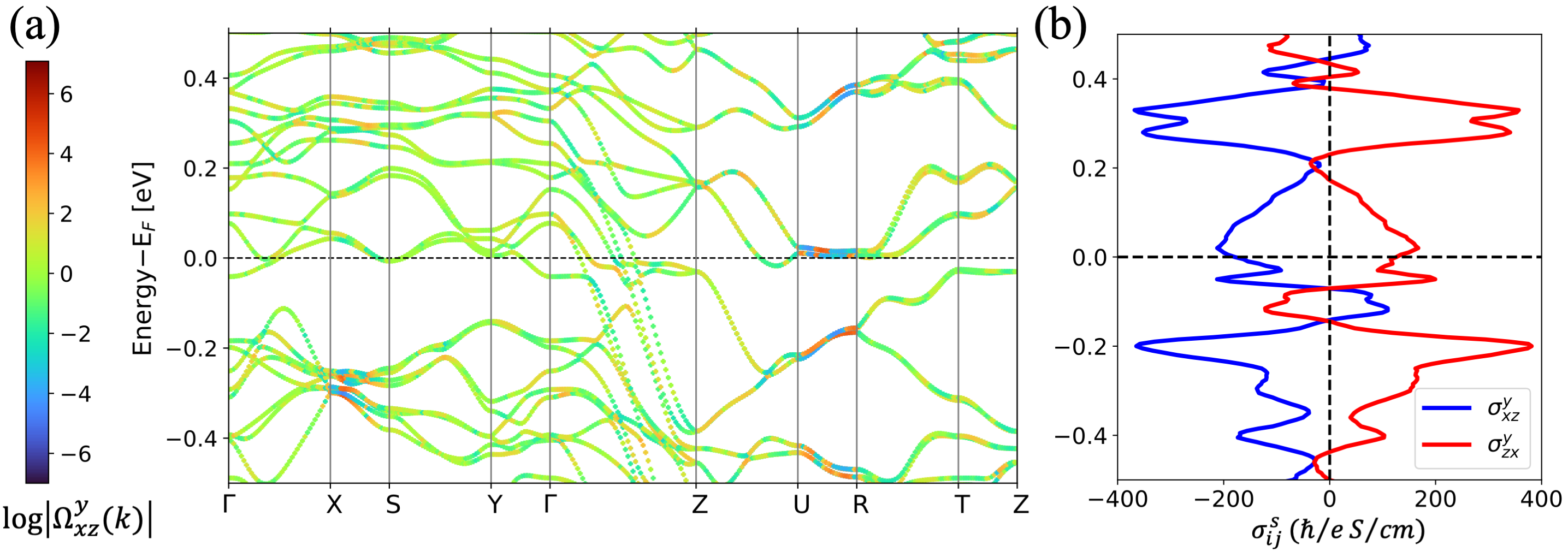}
	\caption{(a) Band structure of Mn$_{5}$Si$_{3}$ in the AF2 phase with spin-orbit coupling. The color scale represents the logarithm of the $k$-resolved spin Berry curvature, $\log \Omega^y_{xz}$, along the Brillouin zone. The SBC intensity is highest at the DNL along the U-R path. This results in a significant signal of the spin Hall conductivity where the DNL is located in energy. (b) The components $\sigma^y_{xz}$ and $\sigma^y_{zx}$ are plotted as a function of energy, with zero energy aligned to the Fermi level.}
	\label{fig3}
\end{figure*}

As mentioned earlier, the band degeneracies along the U-Z path in the SOC case are unaffected because certain symmetries or combinations of them can protect the degeneracies. This demonstrates the stability of this degeneracy near the Fermi level, especially in the presence of SOC. 
Therefore, the collinear AFM Mn$_{5}$Si$_{3}$ possesses DNL even in the presence of SOC, which could have significant effects on the spin transport properties of this compound, making it a promising candidate for studying transport behavior originating from degenerate bands formed by nonsymmorphic symmetries around the Fermi level.

In Figure \ref{fig2}(d), it is also possible to observe the elimination of the degeneracy of the DNL along the U-X-S-R, S-Y-T-R, and Z-T paths in the $k_{\alpha} = \pi$ planes, with $\alpha$ = $x,y,z$, respectively, as illustrated in Figure \ref{fig2}(f). 
Similarly, the eight-fold degenerate DNL along the U-R path splits into two conventionally four-fold degenerate DNLs, a result of the coexistence of the combination $t_{\frac{1}{2},\frac{1}{2},0} \mathbb{T} \widetilde{P}$ and nonsymmorphic symmetries.
These energy gaps generated by the band splitting along these numerous paths in the BZ play a central role in producing a large spin Hall conductivity,  as discussed in the following section.

\section{Spin transport properties}

The $k$-resolved spin Berry curvature (SBC) and spin Hall conductivity (SHC) of the collinear AFM Mn$_{5}$Si$_{3}$ were calculated using the Wannier representation \cite{Pizzi2020}, with results presented in Figure \ref{fig3}. 
In Figure \ref{fig3}(a), a strong signal of the SBC is observed, indicating the presence of a significant SHC, predominantly influenced by the 3$d$-Mn orbital bands near the Fermi level along the U-R path, where the eight-fold degenerate bands split into two sets of four-fold degenerate DNL due to SOC.
Figure \ref{fig3}(b) shows that above the Fermi level, the SHC exhibits a peak close to the Fermi energy, followed by a monotonic decrease until approximately 0.2 eV. 
In contrast, below the Fermi level, a distinct peak is observed at -0.19 eV, which coincides with the energy position of the DNL at the U-R path. 
This indicates that the intrinsic SHC predominantly arises from the DNLs located near the Fermi level, as reflected by the peaks in the SHC as a function of energy.

On the other hand, the SHC calculations are consistent with the symmetry analysis conducted by Seemann $et$ $al$. \cite{Seemann2015} (see Table \ref{tab2}) and show that the collinear AFM Mn$_{5}$Si$_{3}$ exhibits an anisotropic behavior with dominant components $\sigma^{y}_{xz}$, $\sigma^{y}_{zx}$, and $\sigma^{x}_{zy}$. 

\begin{table}[!htb]
	\begin{center}
		\caption{Symmetry analysis for the spin Hall conductivity (SHC) tensor and its numerical results for Mn$_{5}$Si$_{3}$ with SOC. The calculated elements of the SHC tensor are set to zero when they are smaller than 1 ($\hbar/e$) S/cm. This results indicates that the SHC calculations are consistent with the symmetry analysis of the SHC tensor.}
		\label{tab2}
		\vspace{0.05cm}
		\scalebox{1.0}{
			\begin{tabular}{ccccc}
				\hline \hline
				\multicolumn{5}{c}{\textbf{SHC (($\hbar/e$) (S/cm))}}\\
				\hline
				$\underline{\sigma}^{x}$&& $\underline{\sigma}^{y}$ && $\underline{\sigma}^{z}$\\
				\hline
				$\begin{pmatrix}
					0 & 0 & 0\\
					0 & 0 & \sigma^{x}_{yz}\\
					0 & \sigma^{x}_{zy} & 0
				\end{pmatrix}$&& 
				$\begin{pmatrix}
					0 & 0 & \sigma^{y}_{xz}\\
					0 & 0 & 0\\
					\sigma^{y}_{zx} & 0 & 0
				\end{pmatrix}$&& 
				$\begin{pmatrix}
					0 & \sigma^{z}_{xy} & 0\\
					\sigma^{z}_{yx} & 0 & 0\\
					0 & 0 & 0
				\end{pmatrix}$\\
				\hline
				$\begin{pmatrix}
					0 & 0 & 0\\
					0 & 0 & -42\\
					0 & 119 & 0
				\end{pmatrix}$&& $\begin{pmatrix}
					0 & 0 & 126\\
					0 & 0 & 0\\
					-179 & 0 & 0
				\end{pmatrix}$ && 
				$\begin{pmatrix}
					0 & 27 & 0\\
					69 & 0 & 0\\
					0 & 0 & 0
				\end{pmatrix}$\\
				\hline \hline
		\end{tabular}}
	\end{center}
	\vspace{0.03cm}
\end{table}

It is not surprising that ($\sigma^{x}_{zy}$ and $\sigma^{x}_{yz}$), ($\sigma^{y}_{xz}$ and $\sigma^{y}_{zx}$)  and ($\sigma^{z}_{yx}$ and $\sigma^{z}_{xy}$), are not equal in magnitude, as the $x$, $y$, and $z$ directions are not equivalent in this orthorhombic structure. Similar results were reported by Zhang $et$ $al$. \cite{Zhang2017} when studying the SHC of non-collinear AFMs Mn$_{3}$X (X = Ga, Ge, and Sn). Since the peaks of the dominant components $\sigma^{y}_{xz}$ and $\sigma^{y}_{zx}$ shown in Figure \ref{fig3}(b) are located around the Fermi level.
This suggests that employing a doping strategy in Mn$_{5}$Si$_{3}$  could effectively optimize the electronic structure, enhancing the efficiency of spin-to-charge conversion. 
Moreover, the existence of two four-fold degenerate DNLs makes them available to detection through magnetotransport measurements. 
All of the above makes the collinear AFM Mn$_{5}$Si$_{3}$ a highly interesting material for spintronic devices to generate and detect spin currents. 

Furthermore, recently, Reichlova $et$ $al$. \cite{Reichlova2024} conducted theoretical and experimental studies on epitaxial hexagonal layers of Mn$_{5}$Si$_{3}$ grown on a silicon substrate. According to these authors, the altermagnetic AF2 phase exists in a temperature range of 70 K to 240 K and retains the hexagonal structure. This behavior is attributed to considerable strain in the epitaxial layers, causing the $c$ lattice constant to be smaller than the values found in the bulk Mn$_{5}$Si$_{3}$. In this work, it was observed an anomalous Hall conductivity (AHC) with magnitudes ranging from 5 to 20 S cm$^{-1}$, which is attributed to the absence of $t_{\frac{1}{2}} \mathbb{T}$ and $\mathbb{T}$  symmetries. However, in our symmetry analysis, the MSG in this AF2 phase of the orthorhombic Mn$_{5}$Si$_{3}$ compound exhibits the $t_{\frac{1}{2},\frac{1}{2},0} \mathbb{T} \widetilde{P}$ symmetry, which is a combination of $\mathbb{T}$ with inversion and a partial translation. Thus, an expected null AHC is observed \cite{Seemann2015,Reichlova2024}.
Therefore, the AF2 phase of Mn$_{5}$Si$_{3}$ can exhibit a SHC without an accompanying AHE, opening new avenues for the experimental exploration of the SHE to identify the antiferromagnetic orthorhombic phase of Mn$_{5}$Si$_{3}$.

\section{Conclusion}

We conducted an theoretical study of the electronic, magnetic, and transport properties of the orthorhombic compound Mn$_{5}$Si$_{3}$  in the AF2 phase using $ab$-$initio$ calculations and symmetry analysis. 

Our findings reveal that the Mn22 atoms have magnetic moments $\sim$ 2.63 $\mu_{B}$ and $\sim$ 2.26 $\mu_{B}$ with and without SOC respectively, oriented parallel and antiparallel to the crystallographic $b$ axis of the orthorhombic cell. 
The calculated magnetic anisotropy energy indicates that the easy axis of magnetization lies along the [010] direction, consistent with previous experimental observations.

From symmetry analysis, we found that, in the absence of SOC, the observed Dirac nodal planes at $k_{\alpha} = \pi$ ($\alpha$ = $x,y,z$) arise from the symmetries of the spin space group.
Notably, we also uncovered an unconventional eight-fold degenerate Dirac nodal line situated near the Fermi level along the U-R path in the Brillouin zone. This eight-fold degeneracy is protected
by the magnetic space group symmetries that leave the path U-R fixed together with the pure-spin symmetry $ [\widetilde{C_{2y}}||\mathrm{E}] $ and the lattice symmetry
$[\mathrm{E}||t_{0,0,\frac{1}{2}}M_z]$.
When SOC is taken into account, this line splits into two four-fold degenerate DNLs along the U-R path. This energy splitting plays a crucial role in enhancing spin-to-charge conversion via the spin Hall effect.

Collectively, these insights establish collinear AFM Mn$_{5}$Si$_{3}$ as a promising candidate for spintronic devices, especially in the generation and detection of spin currents, while maintaining compatibility with modern silicon technology. 

\section{Appendix}  \label{Appendix}
\subsection{Magnetic-space groups}

Crystals with internal magnetization have symmetry groups that differ from the groups
of symmetries of the crystal alone. The magnetization of an atom in a crystal lattice can be thought
of a direction of magnetization represented by a unit vector $m \in \mathbb{R}^3$. By the spin orbit interaction (SOC), this unit vector is acted upon by the crystal symmetries. 

Let us recall the definition of the magnetic space groups (MSG). The geometrical symmetries
define a subgroup of the group of rigid symmetries of $\mathbb{R}^3$. This group is the semi-direct product 
$ \mathbb{R}^3 \rtimes O(3)$ where the elements are pairs $(t,O)$ with $t \in \mathbb{R}^3$ a translation vector and $O \in O(3)$ an orthonormal matrix, whose group structure is given by the equation
\begin{align}
	(t_1,O_1)(t_2,O_2) = (t_1+O_1t_2,O_1O_2),
\end{align}
and $O(3)$ acts on $\mathbb{R}^3$ by matrix multiplication.

Since we are taking into account the spin of the electrons, the appropriate symmetry group is the 
2-fold cover $\mathbb{R}^3 \rtimes Pin_{-}(3)$ of $\mathbb{R}^3 \rtimes O(3)$, where 
$Pin_{-}(3)$ is the 2-fold cover of $O(3)$ that fits into the following diagram:
\begin{align}
	\xymatrix{
		\{ \pm 1\} \ar[d] & \{ \pm 1 \} \ar[d] & \\
		Spin(3) \ar[r]\ar[d]^{\phi} & Pin_{-}(3) \ar[d]^{\widehat{\phi}}  \ar[dr]^D&\\
		SO(3) \ar[r] & O(3) \ar[r]^{det}& \{\pm 1 \}
	}
\end{align}

The group $Spin(3)$ is nothing else as $SU(2)$, the homomorphism $\phi: SU(2) \to SO(3)$ is the famous double cover

\begin{align} \tiny{
		\left(\begin{matrix} a+ib & c+id\\ - c+ id  & a-ib 
		\end{matrix} \right) \stackrel{\phi}{\mapsto} 
		\left(
		\begin{matrix} 
			1-2c^2-2d^2 & 2bc-2da & 2bd+2ca\\
			2bc+2da & 1-2b^2-2d^2 & 2cd - 2ba \\
			2bd-2ca & 2cd+2ba & 1-2b^2-2c^2 
		\end{matrix}
		\right), }
\end{align}
and the Pin group $Pin_{-}(3)$ is the $\{\pm 1\}$
extension of $O(3)$ where the lifts of mirror symmetries square to $-1$.

The group $Pin_{-}(3)$ is isomorphic to $SU(2) \times \mathbb{Z}_2$, inasmuch as
$O(3)$ is isomorphic to $SO(3) \times \mathbb{Z}_2$. We will denote any element
in $Pin_{-}(3)$ as the product $U \widetilde{P}^a$ where $U \in SU(2)$ and $\widetilde{P}$ is the lift
of the inversion $P$ in $O(3)$. 

The time reversal symmetry (TRS) operator $\mathbb{T}$ is also a symmetry of the system.
In terms of $2 \times 2$ matrices the operator $\mathbb{T}$ is usually defined as $i \sigma_y \mathbb{K}$
where $\sigma_y$ is a Pauli matrix and $\mathbb{K}$ is the complex conjugation operator. 
The TRS symmetry $\mathbb{T}$ commutes with all elements in $Pin_{-}(3)$ and moreover it squares
to $-1$. If we disregard translations, we could take as full symmetry group the group
\begin{align}
	\left( SU(2) \times_{\mathbb{Z}_2} \mathbb{Z}_4 \right) \times \mathbb{Z}_2
\end{align}
where any element can be written as $U \mathbb{T}^a \widetilde{P}^b$, with $U \in SU(2)$,
and keeping in mind that $\mathbb{T}^2 = - \mathrm{Id}$. 

The group of magnetic symmetries is therefore a subgroup of
\begin{align}
	\mathbb{R}^3 \rtimes \left (\left( SU(2) \times_{\mathbb{Z}_2} \mathbb{Z}_4 \right) \times \mathbb{Z}_2 \right)
\end{align}
where the group structure is given by the following equation:
\begin{align}
	(t_1U_1\mathbb{T}^{a_1} & \widetilde{P}^{b_1})(t_2U_2\mathbb{T}^{a_2} \widetilde{P}^{b_2}) =\\ 
	& (t_1 + \phi(U_1)(-1)^{b_1}t_2)U_1U_2 \mathbb{T}^{a_1+a_2} \widetilde{P}^{b_1+b_2}. \nonumber
\end{align}

The MSG is defined as the group of symmetries that leave fixed the position of the atoms in the crystal structure,
and moreover, that leave fixed the internal magnetization. Let us recall how the full symmetry group
acts on the positions of the atoms and their internal magnetization.

Denote by $(m|x) \in    S^2 \times \mathbb{R}^3 $ the position of an atom $x$ and a magnetization unit vector $m$. The magnetization is unaffected by the inversion symmetry $\widetilde{P}$ and the position is unaffected by
the TRS. On the contrary, TRS flips the magnetization and inversion flips the coordinate. The elements
of $SU(2)$ both act by their projection in $SO(3)$ on both the magnetization and the atomic position. Therefore
the action of the full symmetry group is as follows:
\begin{align}
	(tU\mathbb{T}^{a} & \widetilde{P}^{b}) (m|x) = ( \phi(U)(-1)^a m|t+\phi(U)(-1)^bx).
\end{align}

\subsection{Spin-space group}

Decoupling the interaction of the spin and the lattice enlarges the amount of symmetries that a magnetic 
crystal has. These symmetries are called spin-space symmetries (SSG) and are defined as follows.

On the one hand consider the spinorial symmetries. These are the elements of $SU(2)$ plus the TRS.
The elements of this group can be written as $U \mathbb{T}^a$ and they define the group 
$ SU(2) \times_{\mathbb{Z}_2} \mathbb{Z}_4$. Note that $\mathbb{T}$ defines the cyclic group of 4 elements $\mathbb{Z}_4$ but we need that $\mathbb{T}^2$ matches the multiplication by $-1$ in $SU(2)$. They act on the magnetization vector $m$ as follows:
\begin{align} 
	(U \mathbb{T}^a) \cdot m = \phi(U)(-1)^a m.
\end{align}
On the other hand, the symmetries of the lattice $\mathbb{R}^3 \rtimes O(3)$ act only on the position
of the atoms by the following equation:
\begin{align} 
	(tO) \cdot x= t+Ox.
\end{align}
The SSG consists of the symmetries 
\begin{align}
	[U\mathbb{T}^a||tO] \in \left( SU(2) \times_{\mathbb{Z}_2} \mathbb{Z}_4 \right)  \times  \left(\mathbb{R}^3 \rtimes O(3)\right) 
\end{align}
that leave the lattice of atoms with their internal magnetization fixed.

Symmetries that leave the lattice of atoms fixed will be called {\it pure-spin symmetries} and are of the form
$[ U \mathbb{T}^a ||\mathrm{E}]$. Symmetries that leave the magnetization fixed  will be called
{\it lattice symmetries} and are of the form $[\mathrm{E}||tO]$.

The SSG includes the MSG of the magnetic crystal. Any magnetic symmetry can be written separately as
as a composition of a pure-spin symmetry and a lattice symmetry:
\begin{align}
	tU\mathbb{T}^a\widetilde{P}^b \mapsto [ U\mathbb{T}^a||t\phi(U)P^b ].
\end{align}

This assignment defines an injective homomorphism of groups
\begin{align}
	MSG \rightarrowtail  SSG,
\end{align}
where the magnetic symmetries become a subgroup of the spin-space groups symmetries.

Pure-spin symmetries define a the pure-spin group (PSG) which is moreover a normal subgroup of SSG. The quotient
group SSG/PSG is clearly isomorphic to MSG and therefore we conclude that the spin-space group is 
isomorphic to the semi-direct product of the magnetic space group with the pure-spin group:
\begin{align}
	SSG \cong PSG \rtimes MSG,
\end{align}
where the MSG acts on the PSG by conjugation.

\subsection{Pure-spin group}

The structure of the PSG depends on the nature of the magnetization vectors. 
\begin{itemize}
	\item{Colinear magnetization.} Assume the magnetization is parallel to the $z$-axis.
	Then any rotation around the $z$-axis leaves the magnetization fixed. Moreover, TRS composed
	with rotation around the $y$-axis also leaves the magnetization fixed. The group of pure-spin symmetries
	is isomorphic to the group 
	\begin{align}
		U(1) \rtimes \mathbb{Z}_2
	\end{align}
	where $U(1)$ is generated by the lift
	of the rotations of $\theta$ radians around the $z$-axis:
	\begin{align}
		\left(\begin{matrix} \cos(\tfrac{\theta}{2}) & i\sin(\tfrac{\theta}{2})\\ i\sin(\tfrac{\theta}{2})  & \cos(\tfrac{\theta}{2})
		\end{matrix} \right)
	\end{align}
	and the operator $\widetilde{C_{2y}} \mathbb{T}$, which in matrices becomes:
	\begin{align}\widetilde{C_{2y}} \mathbb{T}=
		\left(\begin{matrix} 0 & -1\\ 1  & 0
		\end{matrix} \right) \cdot
		\left(\begin{matrix} 0 & 1\\ -1  & 0
		\end{matrix} \right) \mathbb{K} = \mathbb{K}.
	\end{align}
	Note that the multiplication of $\widetilde{C_{2y}} \mathbb{T}$ on the lift of the rotation matrices reverses
	the direction of the rotation.
	For simplicity one may take the group of pure-spin symmetries generated by the matrices
	\begin{align}
		\widetilde{C_{2z}}, \widetilde{C_{2y}} \mathbb{T}, \widetilde{C_{2x}} \mathbb{T}
	\end{align}
	which is isomorphic the the group $\mathbb{Z}_4 \rtimes \mathbb{Z}_2$. There are only two 1-dimensional  irreducible
	representations such that $(\widetilde{C_{2z}})^2=-1$: $\widetilde{C_{2z}}\psi=\pm i\psi$, $\widetilde{C_{2y}} \mathbb{T} \psi = \mp \psi$.
	\item{Coplanar magnetization.} When the magnetization is coplanar (not colinear) perpendicular to the $z$-axis, 
	the only pure-spin symmetry that the system has is $[\widetilde{C_{2z}}\mathbb{T}||\mathrm{E}]$.
	\item{Non colinear nor coplanar magnetization.} Magnetizations which incorporate vectors not lying on a plane
	do not have pure-spin symmetries. The reason is straightforward; once one fixes two non colinear unitary vectors in $\mathbb{R}^3$, the only rigid transformation leaving them fixed is a mirror or the identity. But mirrors
	are not orientable transformations and therefore do not lie in the image of the map $\phi$.
\end{itemize}

\subsection{Computational methods}

Electronic and magnetic properties of orthorhombic Mn$_{5}$Si$_{3}$ compound in AF2 phase were studied by $ab$-$initio$ calculations based on non-collinear spin density functional theory (DFT). The Kohn-Sham hamiltonian contained scalar-relativistic corrections and SOC was taken into account by a second-variation procedure. The plane-wave projector augmented wave (PAW) method \cite{Blochl1994,Kresse1999} as implemented in vienna ab-initio simulation package (VASP) \cite{Kresse1996,Kresse1996-2} was used. The 3d-Mn electrons were considered valence electrons. In the exchange-correlation functional, we used the generalized gradient approximation (GGA) in the Perdew-Burke-Ernzerhof (PBE) parameterization \cite{Perdew1996}. The electron wave function was expanded in plane waves up to a cutoff energy of 520 eV. A gamma centered grid of 7$\times$4$\times$10 k-points was used to sample the irreducible Brillouin zone in the Monkhorst-Pack special scheme \cite{Monkhorst1976} for the orthorhombic structure, these parameters ensure a convergence better than 1 meV for the total energy. Atomic positions and lattice constants were taken from the experimental data \cite{Brown1992}, and no atomic relaxations were considered. The calculation of the magnetic anisotropy energy was carried out through the self-consistent scheme, which is based on the direct calculation of the total energy difference of the two different magnetization directions with SOC, as shown in the following equation:

\begin{equation}
	MAE = E^{tot}[\hat{m}_{1}]-E^{tot}[\hat{m}_{2}]
\end{equation}
Where $\hat{m}_{1}$ and $\hat{m}_{2}$ are the two different orientations of magnetization \cite{Li2015}.

The intrinsic SHC was evaluated by the Kubo formula \cite{Sinova2015,Guo2008}:
\begin{equation}
	\sigma^{z}_{xy}=\frac{e}{\hbar}\sum_{k}\sum_{n}f_{\textbf{k}n}\Omega^{z}_{n}(\textbf{k}),
\end{equation}
\begin{equation}
	\Omega^{z}_{n}(\textbf{k})=\sum_{n' \neq n}\frac{2Im\left[\langle\textbf{k}n\vert j^{z}_{x}\vert\textbf{k}n'\rangle\langle\textbf{k}n'\vert v_{y}\vert\textbf{k}n\rangle \right]}{\left(\epsilon_{\textbf{k}n} - \epsilon_{\textbf{k}n'} \right)^{2}},
\end{equation}
Where $n'$, $n$ are band indices, $\Omega^{z}_{n}(\textbf{k})$ is the spin Berry curvature for the n-th band, $j^{z}_{x}$ is the spin-current operator, $f_{\textbf{k}n}$ is the Fermi distribution function for the n-th band at \textbf{k}. 
We obtained Wannier functions using the Wannier90 code \cite{Pizzi2020}, and we calculated the spin Hall conductivity by using equation above in Wannier tools \cite{Wu2018}.

\vspace{0.5cm}
\section*{Acknowledgments}
VME acknowledges the financial support of Universidad del Norte and COLCIENCIAS (Administrative Department
of Science, Technology and Research of Colombia) under “Convocatoria 785 - Convocatoria de Doctorados Nacionales 2017". 
RGH gratefully acknowledges the computing time granted on the supercomputer Mogon at Johannes Gutenberg University Mainz (hpc.uni-mainz.de). 
BU acknowledges the support of CONACYT through project CB-2017-2018-A1-S-30345-F-3125 and of the Max Planck Institute for Mathematics in Bonn, Germany. 
RGH and BU thank the continuous support of the Alexander Von Humboldt Foundation, Germany. 
LS acknowledges funding  by the ERC Starting Grant (No. 101165122), Deutsche Forschungsgemeinschaft (DFG, German Research Foundation) 445976410; and TRR288-422213477 (projects A09 and B05).

\bibliographystyle{naturemag}
\bibliographystyle{abbrvnat}
\bibliography{library.bib}

\end{document}